\documentclass[onecollarge]{svjour2}
\usepackage{natbib}
\usepackage{epsfig,graphicx,color,amsmath,subfigure,slashbox,setspace,algorithmic,algorithm,amssymb,cite} 		
\usepackage{color}

\begin{document}
\title{\textbf{Geometrical Domain of Spin-1/2 Probability Mass Function}}

\author{Karthik Bharath\and Swarnamala Sirsi\and A.R.Usha Devi}

\institute{Karthik Bharath\at Department of Systems Science, State University of New York, Binghamton ,PO Box 6000 Binghamton, NY 13902\\
\email{kbharat1@binghamton.edu}
\and Swarnamala Sirsi\at Yuvarajas College, University of Mysore, Mysore 570 005, India\\
\email{ssirsi@uomphysics.net}
\and A.R Usha Devi\at Dept of Physics, Bangalore University, Bangalore, 560 056, India\\
\email{arutth@rediffmail.com}}

\date{ }
\maketitle
\begin{abstract}
 The quantum analogue of the classical characteristic function for a spin 1/2  assembly is considered and  the probability mass function of the random vector associated with the assembly is derived. It is seen that the positive regions of Wigner and Margenau-Hill quasi distributions for the three components of spin, correspond to a trivariate probability mass function. We identify the domain of these positive regions as an Octahedron inscribed in the Bloch sphere with its vertices on the surface of the sphere. It is in this domain that a quantum characteristic function characterizing the quasi distribution, admits a probability mass function in $IR^3$ . It is also observed that the classical variates $X_1$, $X_2$, $X_3$ corresponding to the 3 spin operators $\sigma_1$, $\sigma_2$, $\sigma_3$ in the domain, are independent iff the Bloch vector lies on any one of the axes.
\end{abstract}
\section{INTRODUCTION}
The study of quasiprobability distributions as quantum analogues to phase space distributions in quantum mechanical systems has been well documented~\citep{hill}. These distributions have served as a very helpful tool in analyzing the statistical distributions of spinning submicroscopic particles like electrons, protons etc into various quantum mechanical states and also in providing insights into the connections between classical and quantum mechanics. The Wigner function gives the description of a quantum state in terms of a generalized function which acts a quantum analogue to probability distributions in particle phase space~\citep{wigner}. It was observed that this function assumed negative values in some domains of the phase space  for certain quantum states. The Wigner function is hence a quasiprobability distribution for the position  and momentum of a mechanical system  determining the quantum state $\hat{\rho}$. The  Sudarshan-Glauber's P distribution~\citep{sudarshan}~\citep{glauber}  and the Husimi- Kano's Q distribution~\citep{husimi}~\citep{kano}, adopting a density matrix representation of the quantum state, have attempted to obtain distributions which correspond to positive, normalized classical  distribution functions in phase space.  But the Wigner, Sudarshan-Glauber P distribution and  the Husimi- Kano Q distribution are not \textit{genuine} classical probability distributions owing to the non-commutativity of the quantum mechanical operators which does not permit simultaneous measurement.

More recently, the quantum state reconstruction problem has been addressed using tomographic techniques on the Wigner function\citep{bertrand,ulf,manko1,manko2,ulf2}. This technique uses quadrature histograms, which are radon transformations of the Wigner function. The Wigner function is consequently reconstructed from the histograms using the inverse Radon transformation known as marginal distributions~\citep{vogel}. Tomography has also been used in the measurement of spin states using positive normalized probability distributions wherein tomographic measurements of the density matrix for a quantum state are performed and the the method has been extended to measuring discrete spin operators~\citep{manko4}. The marginal distributions used in tomography play a central role in quantum state tomography and have even been considered as giving a classical-like representation of quantum systems~\citep{manko3}. Here, we explore the classical like nature of the phase space distributions of quantum mechanical systems by considering the Wigner and Margenau-Hill quasiprobability distributions.
 
Let us consider the spin vector $\overrightarrow{S}=(S_1,S_2,S_3)$ of the particle represented by $\frac{\hat{\sigma}}{2}$=$\frac{1}{2}(\sigma_1,\sigma_2,\sigma_3)$ in natural units where $\sigma_1$,$\sigma_2$ and 
$\sigma_3$ are the traditional Pauli's spin matrices satisfying the relations 
\begin{equation}
\label{pauli}
\sigma^2_j=I_2,~\sigma_j\sigma_k=-\sigma_k\sigma_j,~j\not=k=1,2,3
\end{equation}
\[ \sigma_1\sigma_2=i\sigma_3,~\sigma_1\sigma_3=-i\sigma_2,~\sigma_2\sigma_3=i\sigma_1 \]
Each of these three matrices has eigen values $\pm{1}$ and consequently referred to as spin random vectors~\citep{parth}. Given the state $|\Psi \rangle = a| 1/2 \rangle+b|-1/2 \rangle$ of the particle, a classical measurement of the Cartesian components $S_1$,$S_2$ and $S_3$ of spin $\overrightarrow{S}$ is given by the quantum mechanical expectation values 
\begin{equation}
E_{QM}(\overrightarrow{S})=\frac{1}{2}\langle \Psi,\hat{\sigma}\Psi \rangle
\end{equation}
The density matrix representation~\citep{von}~\citep{dirac} of the the quantum mechanical expectation values yields
\begin{equation}
\label{eqn:densitymatrix}
E_{QM}(\overrightarrow{S})=\frac{1}{2}Tr(\rho\hat{\sigma})
\end{equation}
where $Tr$ implies the trace of the density matrix. The density matrix representation allows us to determine the measured value of the spin vector for a statistical assembly, which is given by the expectation value
\begin{equation}
E(\overrightarrow{S})=\frac{1}{N}\sum_{i=1}^{N}\frac{1}{2}Tr(\rho_i\hat{\sigma})
\end{equation}
$E(\overrightarrow{S})$ more specifically denotes the average of the quantum mechanical expectation values for the spin operators of the assembly. $\rho_i$ refers to the individual density matrix associated with the $i_{th}$ particle and N denotes the the total number of particles in the assembly. Significantly, $E(\overrightarrow{S})$ may as well be writted in the form akin to equation (\ref{eqn:densitymatrix}) as 
\begin{equation}
\label{eqn:densitymatrix1}
E(\overrightarrow{S})=\frac{1}{2}Tr(\rho\hat{\sigma})
\end{equation}
where $\rho=\frac{1}{N}\sum_{i=1}^{N}\rho_i$ is referred to as the density matrix for the statistical assembly of spinning particles.

It is but natural to ask the question 'Can we identify these expectation values as moments of some statistical distribution?'. What we are looking for is a trivariate probability mass function $P(X_1=x_1,X_2=x_2,X_3=x_3)$ of classical random variables $X_1,X_2,X_3$ corresponding to the three spin components, which must be able to reproduce exactly the expectation value $E(S_1,S_2,S_3)$  as given by equation (\ref{eqn:densitymatrix1}). It is important, in this context, to keep in mind that $S_1,S_2,S_3$ are typically non commuting quantum mechanical observables. We identify the characteristic function of the random variables $X_1,X_2,X_3$  which corresponds to the spin operators at the expectation value level. In other words, the three classical random variables correspond to the three spin operators in the sense that they produce the same expectation values. Further, it has been observed that in the pure state $|e_1 \rangle = \left( \begin{array}{c} 
1 \\ 
0 \end{array} \right)$ 
\begin{equation}
\phi(t_1,t_2) = \langle e_1,e^{(it_1\sigma_1+it_2\sigma_2)}e_1 \rangle
\end{equation}
is not a characteristic function of any probability distribution in the plane~\citep{parth1}. Parthasarathy also mentions the inconclusiveness of the conditions on a unit vector $|u \rangle$ and a family of observables $X_1,X_2\ldots.X_k$ in a Hilbert space $\cal{H}$, the function $\phi(t_1,t_2,\ldots,t_k) = \langle u,e^{(it_1X_1+it_2X_2+\ldots+it_kX_k)}u \rangle$ is in $IR^k$. In this paper, we look to identify the conditions and consider the quantum analogue of the classical characteristic function $(E(e^{(it_1X_1+it_2X_2+it_3X_3)})$ ~\citep{chan} for a spin half assembly and identify the domain in which it admits a trivariate probability mass function in $IR^3$. In section 3 we take into consideration the interesting Bloch vector $\overrightarrow{p} =(p_1,p_2,p_3)$ and parameterize the density matrix using the Bloch vector and state the conditions for pure and mixed states ensembles. In section 4, we briefly discuss the Wigner-Weyl and Margenau-Hill quasiprobability distribution functions  and the rules which associate a classical characteristic function to the corresponding quantum mechanical operator and identify the geometrical domain in which the Margenau-Hill characteristic function admits a probability mass function for the quantum spin $1/2$ assembly.

\section{SPIN -$1/2$ DENSITY MATRIX}
For a mixture state the density matrix $\rho$ is given by
\begin{equation}
\rho = \sum_{k=1}^{N}w_k|\Psi_k\rangle\langle\Psi_k|
\end{equation}
where $w_k=N_k/N$ is the probability of finding the particle in the state $a_k| 1/2 \rangle + b_k|-1/2 \rangle$ and $N_k$ stands for the number of particles in the state $|\Psi_k \rangle$. Since the density matrix $\rho$ is Hermitian and $Tr(\rho)$ is normalized to unity, only three real parameters and hence three measurements are necessary to specify $\rho$ in its entirety. It is customary to represent the density matrix of a spin-$1/2$ system by
\begin{equation*}
\rho=\frac{1}{2}(I_2+\sigma_1p_1+\sigma_2p_2+\sigma_3p_3)
\end{equation*}
\begin{equation}
\label{density}
    =  \frac{1}{2}      \left( \begin{array}{cc}
                          1+p_3 & p_1-ip_2\\
                          p_1+ip_2 & 1-p_3 \end{array} \right)
\end{equation}

The Bloch vector $\overrightarrow{p}$ is given by
\begin{equation}
\label{polarization}
p_k = Tr(\rho\sigma_k) = 2E_{QM}(S_k)
\end{equation}

Since the density matrix is positive semidefinite with the diagonal entries representing probabilities, we have for mixed states
\begin{equation}
\label{sphere}
p_1^2+p_2^2+p_3^2 \leq 1
\end{equation}

Since the density operator for pure states is idempotent, it can be  easily verified that
\begin{equation}
Tr(\rho^2) = (Tr\rho)^2 = 1
\end{equation}

in other words,
\begin{equation}
p_1^2+p_2^2+p_3^2 = 1
\end{equation}

Clearly, for mixture states $Tr({\rho}^2)<1$ and consequently the state of polarization of a spin 1/2 assembly may be represented by a point within the sphere
\begin{equation}
\label{sphere1}
p_1^2+p_2^2+p_3^2 < 1
\end{equation} 

\section{DOMAIN OF DISTRIBUTION FOR SPIN-1/2 ASSEMBLY}

The correspondence between an observable in quantum probability and a random variable in classical probability has been  recognized~\citep{parth1}. These observables in general, are non-commuting and represented by Hermitian matrices. Of the various correspondence rules which associate classical quantities to quantum mechanical operators~\citep{cohen}, the Wigner-Weyl correspondence rule~\citep{hill} characterizing the Wigner quasiprobability distribution can be stated as
\begin{equation}
e^{(At_1+Bt_2)}\rightarrow e^{(\textbf{A}t_1+\textbf{B}t_2)}
\end{equation} 

or

\begin{equation}
A^mB^n\rightarrow \{\textbf{A}^m\textbf{B}^n\}
\end{equation}

where A and B represent the classical variates and \textbf{A} and \textbf{B} represent non-commuting quantum mechanical operators. The \{ \} stands for the total symmetrizer with $(m+n)!/m!n!$ terms. It can be noted that $\phi(t_1,t_2) = \langle e_1,e^{i(\sigma_1t_1+\sigma_2t_12)}e_1\rangle$ considered in~\citep{parth1}, follows the Wigner-Weyl association rule for a bivariate distribution. In a trivariate case for a more general $\rho$, $\phi(t_1,t_2,t_3)$ can be written as 

\begin{equation}
\phi(t_1,t_2,t_3) = Tr(\rho e^{i(\sigma_1t_1+\sigma_2t_2+\sigma_3t_3))} \\
                           = \cos(\|\overrightarrow{t}\|)+i\frac{\langle \overrightarrow{p},\overrightarrow{t} \rangle}{\|\overrightarrow{t}\|} \sin(\|\overrightarrow{t}\|)
\end{equation}
where $\|\overrightarrow{t}\|$ denotes the norm of the vector $\overrightarrow{t} = (t_1,t_2,t_3)$.\\
\\
The Margenau-Hill~\citep{hill} symmetrization rule can be written in the form

\begin{equation}
\label{margenau}
e^{i(At_1+Bt_2)}\rightarrow \frac{1}{2}(e^{i\textbf{At}_1}e^{i\textbf{B}t_2}+e^{i\textbf{Bt}_2}e^{i\textbf{A}t_1})
\end{equation}

or equivalently
\begin{equation}
A^mB^n\rightarrow \frac{1}{2}(\textbf{A}^m\textbf{B}^n + \textbf{B}^n\textbf{A}^m)
\end{equation}

It is to be noted that the moments obtained from the Margenau-Hill and the Wigner-Weyl distributions are the same for a spin $1/2$ assembly. We hence consider the Margenau-Hill distribution and identify the domain in which the quantum characteristic function for the distribution admits a trivariate probability mass function in $IR^3$.

Generalizing equation (\ref{margenau}) for three spin operators, the characteristic function is defined as~\citep{chan}

\begin{equation}
\label{bigone}
\phi(t_1,t_2,t_3) = \frac{1}{3!}(\rho(\lambda_{123}+\lambda_{132}+\lambda_{213}+\lambda_{231}+\lambda_{312}+\lambda_{321}))
\end{equation}

where
\begin{equation}
\lambda_{abc} = e^{i\sigma_at_a}e^{i\sigma_bt_b}e^{i\sigma_ct_c}
\end{equation}

Using equations (\ref{pauli}),(\ref{density}),(\ref{polarization}) together with the relations

\begin{equation*}
e^{i\sigma_kt_k} = I_2\cos(t_k)+i\sigma_k \sin(t_k);k = 1,2,3
\end{equation*}
\begin{equation}
\langle\hat{\sigma},\overrightarrow{A}\rangle \langle\hat{\sigma},\overrightarrow{B}\rangle = \langle\overrightarrow{A},\overrightarrow{B}\rangle + i\langle\hat{\sigma},(\overrightarrow{A}\times\overrightarrow{B})\rangle
\end{equation}

where $(\times)$ refers to the vector cross product, we can simplify equation (\ref{bigone}), which yields

\begin{eqnarray}
\nonumber \phi(t_1,t_2,t_3) = \cos(t_1) \cos(t_2) \cos(t_3) + ip_1 \sin(t_1) \cos(t_2) \cos(t_3)\\
+ip_2 \cos(t_1)\sin(t_2) \cos(t_3) +ip_3 \cos(t_1) \cos(t_2) \sin(t_3)
\end{eqnarray}

Upon inverting the above equation and obtaining the three independent moments for the trivariate distribution, we find that it is the characteristic equation of a discrete trivariate random vector $(X_1,X_2,X_3)$ with a probability mass function 

\begin{equation}
 P(X_1=x_1,X_2=x_2,X_3=x_3) = \frac{1}{8}(1+x_1p_1+x_2p_2+x_3p_3)\\
\end{equation}\\

\noindent with
\begin{equation*}
x_1,x_2,x_3=\pm1
\end{equation*}
provided
\begin{equation*}
\vert p_1\pm p_2\pm p_3\vert\leq1
\end{equation*}
and
\begin{equation}
p_1^2+p_2^2+p_3^2 \leq 1
\end{equation}
 
\begin{figure}[t]
\includegraphics{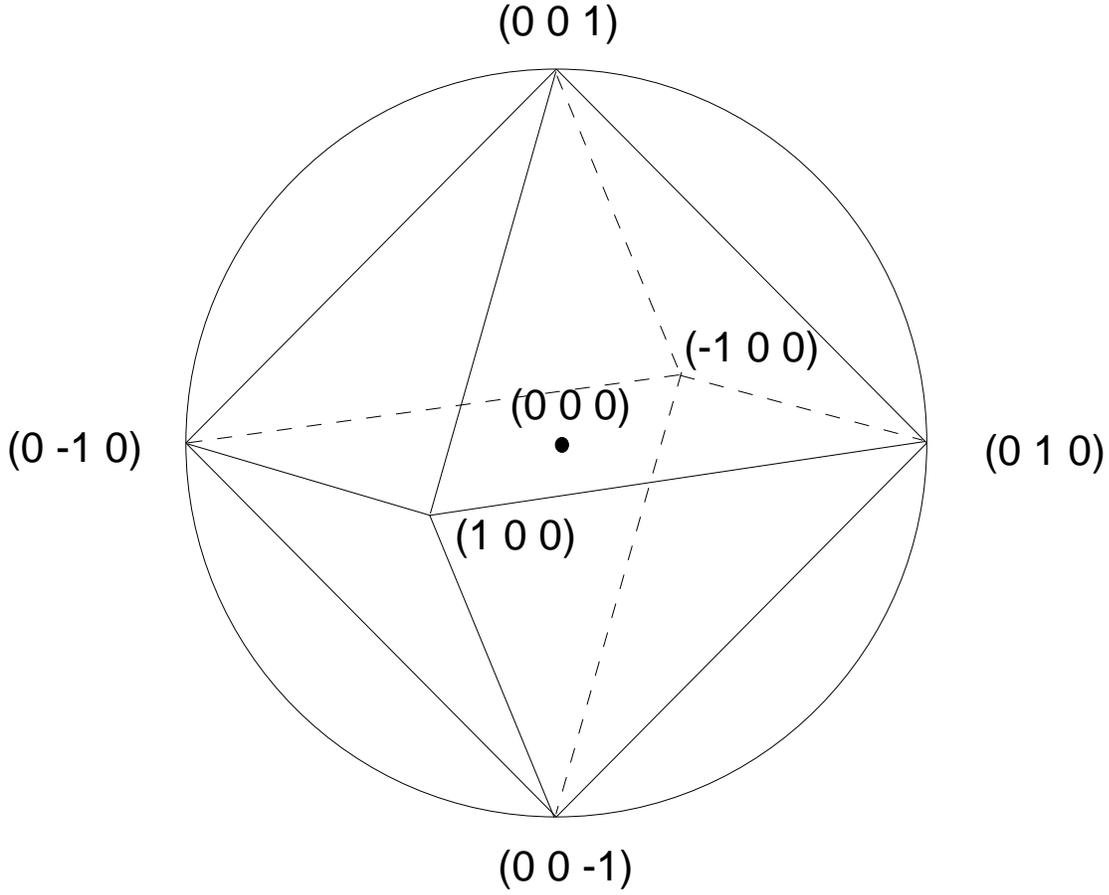}
\label{Figure1}
\caption{Pictorial representation of the geometrical domain in which the Wigner and Margenau-Hill quasiprobability distributions take positive values in $IR^3$. The octahedron inscribed in the sphere, $IR^3$, is the geometrical domain where the quantum characteristic function of the quasiprobability distribution admits a classical trivariate probability mass function. The random variables $X_1,X_2,X_3$ of the probability mass function produce expectation values which correspond to the expectation values produced by the 3 spin operators $S_1,S_2,S_3$, inside the octahedron. The vertices of the octahedron, on the surface of the sphere, represent the pure states of the statistical assembly.}
\end{figure} 

We now have the geometrical domain in which the quantum characteristic function of a quasiprobability distribution for the $3$ spin operators, admits a probability mass function. Geometrically, $\phi(t_1,t_2,t_3)$ is a characteristic function iff the Bloch vector $\overrightarrow{p}$ lies inside an octahedron inscribed within the sphere given by equation (\ref{sphere1}), with its center coinciding with the center of the sphere and its vertices touching the sphere at the points $(\pm1,0,0),(0,\pm1,0),(0,0,\pm1)$ as shown in Fig.1. It in understood that the vertices of the octahedron represent pure states. It is to be noted that the classical variates $X_1,X_2$ and $X_3$ are generally not independent. They are independent iff any two of the components of the Bloch vector $\overrightarrow{p}$, $p_k$'s, are equal to zero. In other words,  $X_1,X_2$ and $X_3$ are independent iff $\overrightarrow{p}$ lies on one of the axes. Further, it can be seen that if $p_k=0$ for some $k(k=1,2,3)$, then $X_k$ is independent of each of the other two random variables.  
 

\section{CONCLUDING REMARKS}

We have identified  the geometrical domain where a classical trivariate probability mass function is admitted by the quantum characteristic function of a  quasiprobability distribution of the three spin operators, in $IR^3$.  The octahedron inscribed in the sphere signifies the domain in which the Wigner and Margenau-Hill quasidistributions assume positive regions and produce expectation values which correspond to the expectation values produced by the normalized, positive classical trivariate probability mass function. It is to be noted that the  three random variables of the probability mass function do not imply commutativity. The correspondence between the three noncommuting spin operators and the three random variables of the classical probability mass function is \textit{only} at the expectation value level.


\bibliography{references}

\begin{thebibliography}{10}

\bibitem{hill}
M.~Hillery, R.F. O'Connel, M.O. Scully, and E.P. Wigner.
\newblock {Distribution Functions in Physics}.
\newblock {\em {Physics Reports}}, 106:123--167, 1984.

\bibitem{wigner}
E.~Wigner.
\newblock {On the Quantum Correction For Thermodynamic Equilibrium}.
\newblock {\em {Physical Review}}, 40:749--759, 1932.

\bibitem{sudarshan}
E.C.G Sudarshan.
\newblock {Equivalence of Semiclassical and Quantum Mechanical Descriptions of
  Statistical Light Beams }.
\newblock {\em {Physical Review Letters}}, 10:277--279, 1963.

\bibitem{glauber}
R.J. Glauber.
\newblock {Photon Correlations}.
\newblock {\em {Physical Review Letters}}, 10:84--86, 1963.

\bibitem{husimi}
K.~Husimi.
\newblock {\em {Proc. Phys. Math. Soc. Japan}}, 22:264, 1940.

\bibitem{kano}
Y.~Kano.
\newblock {\em {Journal of Mathematical Physics}}, page 1913, 1965.

\bibitem{bertrand}
J.~Bertrand and P.~Bertrand.
\newblock {A Tomographic Approach to Wigner Function}.
\newblock {\em {Foundations of Physics}}, 17:397--405, 1987.

\bibitem{ulf}
U.~Leonhardt.
\newblock {Quantum State Tomography and Discrete Wigner Function}.
\newblock {\em {Physical Review Letters}}, 74:4101--4105, 1995.

\bibitem{manko1}
S.~Mancini, V.I. Man'ko, and P.~Tombesi.
\newblock {Symplectic tomography as classical approach to quantum systems}.
\newblock {\em {Phys.Lett.A}}, 213:1, 1996.

\bibitem{manko2}
S.~Mancini, V.I. Man'ko, and P.~Tombesi.
\newblock {Different Realizations of the Tomographic Principle in Quantum State
  Measurement}.
\newblock {\em {Journal of Modern Optics}}, 44:2281, 1997.

\bibitem{ulf2}
U.~Leonhardt.
\newblock Measuring the quantum state of light.
\newblock {\em Measurement Science and Technology}, 11(12):1827--1828, 2000.

\bibitem{vogel}
K.~Vogel and H.~Risken.
\newblock {Determination of quasiprobability distributions in terms of
  probability distributions for the rotated quadrature phase }.
\newblock {\em {Physical Review A}}, 40:2847--2849, 1989.

\bibitem{manko4}
V.I Man'ko and O.V. Man'ko.
\newblock {Spin State Tomography }.
\newblock {\em {J. Exp. Theor. Phys.}}, 85:430, 1997.

\bibitem{manko3}
M.~Caponigro, S.~Mancini, and V.I. Man'ko.
\newblock {A probabilistic approach to quantum mechanics based on 'tomograms'
  }.
\newblock {\em {Fortschritte der Physik}}, 54:602--612, 2006.

\bibitem{parth}
K.R. Parthasarathy.
\newblock {A Remark on Spin Correlations}.
\newblock {\em {Sankhya}}, 51A:192, 1989.

\bibitem{von}
Von Neuman.
\newblock {Thermodynamik Quanten Mechaniser Gesamtheiten}.
\newblock {\em {Nachr.Ges.Wiss Gottingen}}, 245:273, 1927.

\bibitem{dirac}
P.A.M. Dirac.
\newblock The basis of statistical quantum mechanics.
\newblock In {\em Proc.Cambridge Phil.soc}, volume~25, page~62, 1929.

\bibitem{parth1}
K.R. Parthasarathy.
\newblock {ISI Lectures on Quantum Stochastic Calculus}.
\newblock {\em {Sankhya}}, 1988.

\bibitem{chan}
C.~Chandler, L.~Cohen, C.~Lee, M.O. Scully, and K.~Wodkiewiez.
\newblock {Quasi-probability Distribution for Spin-1/2 particle}.
\newblock {\em {Foundations of Physics}}, 22:867, 1992.

\bibitem{cohen}
L.~Cohen.
\newblock {Generalised Phase Space Distribution Functions }.
\newblock {\em {Journal of Mathematical Physics}}, 7:781, 1966.

\end{thebibliography}

\bibliographystyle{unsrt}
\end{document}